\documentstyle[multicol,aps,psfig]{revtex}
\renewcommand{\narrowtext}{\begin{multicols}{2}
\global\columnwidth20.5pc\noindent}
\renewcommand{\widetext}{\end{multicols}
\global\columnwidth42.5pc}
\begin{document}
\draft
\preprint{16 May 2002}
\title{Spin-Wave Description of Nuclear Spin-Lattice Relaxation in
       Mn$_{12}$O$_{12}$ Acetate}
\author{Shoji Yamamoto and Takashi Nakanishi}
\address{Division of Physics, Hokkaido University,
         Sapporo 060-0810, Japan}
\date{16 May 2002}
\maketitle
\begin{abstract}
In response to recent nuclear-magnetic-resonance (NMR) measurements
on the molecular cluster Mn$_{12}$O$_{12}$ acetate, we study the
nuclear spin-lattice relaxation rate $1/T_1$ developing a modified
spin-wave theory.
Our microscopic new approach, which is distinct from previous
macroscopic treatments of the cluster as a rigid spin of $S=10$, not
only excellently interprets the observed temperature and
applied-field dependences of $1/T_1$ for $^{55}$Mn nuclei but also
strongly supports the $^{13}$C NMR evidence for spin delocalization
over the entire molecule.
\end{abstract}
\narrowtext

\begin{figure}
\centerline
{\mbox{\psfig{figure=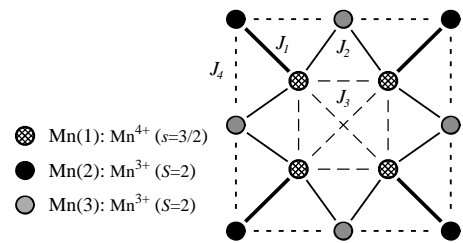,width=60mm,angle=0}}}
\vspace*{4mm}
\caption{Schematic plot of the Mn12 cluster.
         Inequivalent sites Mn(1), Mn(2), and Mn(3) are occupied by
         Mn$^{4+}$ ions with $s=\frac{3}{2}$, Mn$^{3+}$ ions with
         $S=2$, and Mn$^{3+}$ ions with $S=2$, respectively.
         There exist four types of exchange interaction between them:
         $J_1$, $J_2$, $J_3$, and $J_4$, which are drawn by thick
         solid, dashed, thin solid, and dotted lines, respectively.}
\label{F:illust}
\end{figure}

   Mesoscopic magnetism \cite{G1054} is one of the hot topics in
materials science, where we can observe a quantum-to-classical
crossover on the way from molecular to bulk magnets.
Metal-ion magnetic clusters are thus interesting and among others is
[Mn$_{12}$O$_{12}$(CH$_3$COO)$_{16}$(H$_2$O)$_4$] \cite{L2042}
(hereafter abbreviated as Mn12), for which quantum tunneling of the
magnetization \cite{F3830,T145} was observed for the first time.
There are three symmetry-inequivalent Mn sites in the Mn12 cluster
(see Fig. \ref{F:illust}).
The four inner Mn$^{4+}$ spins and the eight outer Mn$^{3+}$ spins
are directed antiparallel to each other and exhibit a novel ground
state of total spin $S=10$ \cite{C5873}.
Resonant magnetization tunneling in such high-spin molecules can be
an evidence for the validity of quantum mechanical approaches at the
nanometer scale.

   The simplest Hamiltonian for the Mn12 cluster in a field may be
given by
$
   {\cal H}
   =-D(S^z)^2
    -g\mu_{\rm B}\mbox{\boldmath$S$}\cdot\mbox{\boldmath$H$}\,,
$
where the molecular cluster is strictly treated as a rigid $S=10$
object with single-axis magnetic anisotropy.
While such a macroscopic treatment of the molecule interprets well
quantum relaxation of the magnetization \cite{P5794,M395}, recent
electron-paramagnetic-resonance (EPR) measurements \cite{H2453}
suggest a possible breakdown of the spin-$10$ description.
Although a microscopic treatment of each Mn moment is necessary
for further understanding of nanoscale magnets and is interesting in
itself, the total spin states in the Mn12 cluster is too large even
for modern computers to directly handle.
Thus it is an idea \cite{S1804} that the Mn$^{3+}$-Mn$^{4+}$ pairs
connected by the strongest exchange interaction $J_1$ construct
composite spins $\frac{1}{2}$.
This eight-spin scheme explains well the magnetization
\cite{F3830,Z1140}, inelastic-neutron-scattering \cite{H8819,K6919},
and EPR \cite{K6919,B8192} measurements at sufficiently low
temperatures.
However, a recent skillful numerical-diagonalization study
\cite{R064419} has shown that even the ground state is quite
sensitive to the less dominant exchange interactions $J_2$, $J_3$,
and $J_4$ as well, throwing doubt on a parameter assignment within
the eight-spin scheme \cite{S1804}, $J_1\sim -150\,\mbox{cm}^{-1}$,
$J_2\simeq J_3\sim -60\,\mbox{cm}^{-1}$, and $J_4\sim 0$.

   It is also unfortunate for this eight-spin model to be less
applicable to nuclear-magnetic-resonance (NMR) measurements, which
have vigorously been carried out for the Mn12 cluster
\cite{L514,A2941,F14246,G1227,F104401,A064420} in recent years.
Although the nuclear spin-lattice relaxation time $T_1$ can serve
as a probe to the electron spin dynamics within the molecule, it is
not yet interpreted beyond the spin-$10$ description.
The early NMR studies on the Mn12 cluster were performed by using
proton \cite{L514,F14246} or deuteron \cite{A2941} nuclei as probes
and therefore the interpretation was plagued by the averaging effect
over the numerous protons and the weakness of the hyperfine coupling
to the Mn moments.
In order to obtain more direct information on the local magnetic
properties, $^{55}$Mn NMR measurements \cite{G1227,F104401} have
recently been performed.
There has also appeared an elaborate $^{13}$C NMR evidence
\cite{A064420} of the paramagnetic spin density of the cluster being
delocalized over the entire molecule.
Now we make our first attempt to interpret microscopically the
nuclear spin-lattice relaxation-time measurements, developing a
modified spin-wave theory \cite{Y14008}.

   We introduce a microscopic Hamiltonian for the Mn12 cluster as
\begin{eqnarray}
   &&
   {\cal H}
   =-\sum_{l=1}^N
    \Bigl[
     2J_1\mbox{\boldmath$s$}_{l}\cdot\mbox{\boldmath$S$}_{l}
    +2J_2(\mbox{\boldmath$s$}_{l}\cdot
          \mbox{\boldmath$\widetilde{S}$}_{l}
         +\mbox{\boldmath$\widetilde{S}$}_{l}\cdot
          \mbox{\boldmath$s$}_{l+1})
   \nonumber \\
   &&
    +2J_3(\mbox{\boldmath$s$}_{l}\cdot\mbox{\boldmath$s$}_{l+1}
         +\frac{1}{2}
          \mbox{\boldmath$s$}_{l}\cdot\mbox{\boldmath$s$}_{l+2})
    +2J_4(\mbox{\boldmath$S$}_{l}\cdot
          \mbox{\boldmath$\widetilde{S}$}_{l}
         +\mbox{\boldmath$\widetilde{S}$}_{l}\cdot
          \mbox{\boldmath$S$}_{l+1})
   \nonumber \\
   &&
    +D_2(S_{l}^z)^2
    +D_3(\widetilde{S}_{l}^z)^2
    +g\mu_{\rm B}H(s_{l}^z+S_{l}^z+\widetilde{S}_{l}^z)
    \Bigr]\,,
   \label{E:H}
\end{eqnarray}
\noindent
where $\mbox{\boldmath$s$}_l$, $\mbox{\boldmath$S$}_l$, and
$\widetilde{\mbox{\boldmath$S$}}_l$ are the spin operators for the
Mn(1) (spin $\frac{3}{2}\equiv s$), Mn(2) (spin $2\equiv S$), and
Mn(3) (spin $2\equiv S$) sites in the $l$th unit, and $N$, the number
of the elementary units, is equal to $4$.
The strongest exchange interaction $J_1$ and the next leading
antiferromagnetic interaction $J_2$ were reliably estimated at
$-150\,\mbox{cm}^{-1}$ and $-60\,\mbox{cm}^{-1}$, respectively, but
the rest of the parameters remain to be fixed.
Taking account of previous investigations, we adopt two contrasting
parameter sets:
(a) $J_1=-150\,\mbox{cm}^{-1},\,J_2=-60\,\mbox{cm}^{-1},\,
     J_3=  60\,\mbox{cm}^{-1},\,J_4=30\,\mbox{cm}^{-1}$
    \cite{Z1140,C938};
(b) $J_1=-150\,\mbox{cm}^{-1},\,J_2=-60\,\mbox{cm}^{-1},\,
     J_3= -30\,\mbox{cm}^{-1},\,J_4=30\,\mbox{cm}^{-1}$
    \cite{S1804,R064419}.
As for the anisotropy parameters, there is much less information.
When the molecule is treated as a rigid spin-$10$ object, the
macroscopic uniaxial crystalline anisotropy parameter $D$ is
determined so as to fit the zero-field separation between the
$M=\pm 10$ and $M=\pm 9$ levels, which is about $14\,\mbox{K}$
\cite{C5873,S1804}.
Hence it is natural to choose the local single-ion anisotropy
parameters, which describe the Jahn-Teller-distorted Mn$^{3+}$ ions
\cite{S141}, within the same scheme \cite{K6919}.
Setting $D_2$ and $D_3$ both equal to $1.5\,\mbox{cm}^{-1}$, we
obtain the excitation energy of $13.7\,\mbox{K}$ in the following
spin-wave treatment.
The $g$ factors are all set equal to 2 \cite{F104401}.

\begin{figure}
\centerline
{\mbox{\psfig{figure=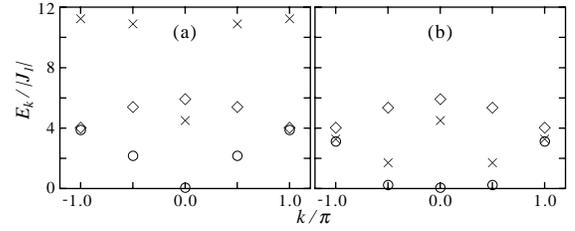,width=74mm,angle=0}}}
\caption{Dispersion relations of the ferromagnetic ($\circ,\diamond$)
         and antiferromagnetic ($\times$) spin waves, which lie in
         the subspace of $M=9$ and that of $M=11$, respectively, for
         the parameter sets (a) and (b).}
\label{F:dsp}
\end{figure}

   We consider a spin-wave treatment of the Hamiltonian (\ref{E:H})
introducing the bosonic operators for the spin deviation in each
sublattice via
$s_l^z=-s+a_{l,1}^\dagger a_{l,1}$,
$s_l^+=\sqrt{2s}a_{l,1}^\dagger$;
$S_l^z=S-a_{l,2}^\dagger a_{l,2}$,
$S_l^+=\sqrt{2S}a_{l,2}$;
$\widetilde{S}_l^z=S-a_{l,3}^\dagger a_{l,3}$,
$\widetilde{S}_l^+=\sqrt{2S}a_{l,3}$.
We carry out the Bogoliubov transformation in the momentum space,
\begin{equation}
   \left.
   \begin{array}{rrrr}
    a_{k,1}=&-\psi_{11}  (k)\,b_{k,1}^\dagger
            &-\psi_{12}  (k)\,b_{k,2}^\dagger
            &+\psi_{13}  (k)\,b_{k,3}\,, \\
    a_{k,2}=& \psi_{21}^*(k)\,b_{k,1}
            &+\psi_{22}^*(k)\,b_{k,2}
            &-\psi_{23}^*(k)\,b_{k,3}^\dagger\,, \\
    a_{k,3}=& \psi_{31}^*(k)\,b_{k,1}
            &+\psi_{32}^*(k)\,b_{k,2}
            &-\psi_{33}^*(k)\,b_{k,3}^\dagger\,, \\
   \end{array}
   \right.
\end{equation}
so as to reach the diagonal Hamiltonian
\begin{equation}
   {\cal H}
   =E_{\rm g}
   +\sum_k\sum_{j=1,2,3}\omega_j(k)\,b_{k,j}^\dagger b_{k,j}\,,
\end{equation}
where 
$E_{\rm g}=8Ss(J_1+2J_2)-12s^2J_3-16S^2J_4-4S^2(D_2+D_3)
     -g\mu_{\rm B}H(2S-s)$.
The numerically calculated dispersion relations $\omega_i(k)$ for the
two parameter sets are shown in Fig. \ref{F:dsp}.
The lowest-lying ferromagnetic ($\circ$) and the antiferromagnetic
($\times$) branches are both sensitive to $J_3$, while the second
ferromagnetic ($\diamond$) branch exhibits little dependence on
$J_3$.

   The core idea of the so-called modified spin-wave theory is
summarized as constructing reliable thermodynamics in low dimensions
by controlling the boson number.
Constraining the total magnetization to be zero, Takahashi
\cite{T168} obtained an excellent description of the low-temperature
thermodynamics of one-dimensional Heisenberg ferromagnets.
His idea that the thermal spin deviation, that is, the number of
thermally induced bosons, should be equal to the ground-state
magnetization may be replaced by \cite{Y14008}
\vspace*{0mm}
\begin{equation}
   \sum_k
   \sum_{j=1,2,3}
    \bar{n}_{k,j}
   \sum_{i=1,2,3}
    |\psi_{ij}(k)|^2
   =8S+4s\,,
   \label{E:const}
\end{equation}
for the ferrimagnetic Mn12 cluster, where $\bar{n}_{k,j}$ is
expressed as $\sum_{n_1,n_2,n_3=0}^\infty n_j P_k(n_1,n_2,n_3)$ with
$P_k$ being the probability of $n_j$ spin waves of mode-$j$ appearing
in the $k$-momentum state.
Equation (\ref{E:const}) claims that the thermal fluctuation should
cancel the {\it staggered} magnetization instead of the {\it uniform}
one, in response to the present ferrimagnetic ground state.
Minimizing the free energy
$F=E_{\rm g}+\sum_k\sum_{j=1,2,3}\bar{n}_{k,j}\omega_j(k)
  +k_{\rm B}T\sum_k\sum_{n_1,n_2,n_3}
   P_k(n_1,n_2,n_3){\rm ln}P_k(n_1,n_2,n_3)$
with respect to $P_k$ at each $k$ under the condition (\ref{E:const})
together with the trivial constraints
$\sum_{n_1,n_2,n_3}P_k(n_1,n_2,n_3)=1$, we obtain the optimum
distribution functions as
\begin{equation}
   \bar{n}_{k,j}
   =\frac{1}
    {{\rm e}^{[\omega_j(k)+\mu\sum_{i=1,2,3}
                              |\psi_{ij}(k)|^2]/k_{\rm B}T}
   -1}\,,
\end{equation}
where $\mu$ is a Lagrange multiplier due to Eq. (\ref{E:const}).
Once the distribution is determined, we can readily calculate any
thermal quantities such as the internal energy
$U=E_{\rm g}+\sum_k\sum_{j=1,2,3}\bar{n}_{k,j}\omega_j(k)$ and
the magnetic susceptibility
$\chi=[(g\mu_{\rm B})^2/k_{\rm B}T]
      \sum_{j=1,2,3}\bar{n}_{k,j}(1+\bar{n}_{k,j})$.

   Although the above-demonstrated modified spin-wave scheme of the
ferrimagnetic version generally works well in low dimensions
\cite{Y11033,N}, it is not yet applicable to the Mn12 cluster as it
is.
Due to the significant Jahn-Teller distortion, the lowest excited
states of $M=\pm 9$ are separated from the ground states of
$M=\pm 10$ by a finite energy $\mit\Delta$, which is incompatible
with the condition (\ref{E:const}).
In isotropic ferrimagnets, there exists a zero-energy excitation and
therefore a certain number of bosons naturally survive at low
temperatures.
The grandcanonical constraint (\ref{E:const}) not only works so as
to suppress the thermal divergence of the boson number at high
temperatures but also gives a precise description of the
low-temperature thermodynamics \cite{Y11033}.
On the other hand, once a gap $\mit\Delta$ opens, the boson number
should exponentially decreases as
$\propto{\rm e}^{-{\mit\Delta}/k_{\rm B}T}$ at low temperatures,
while the constraint (\ref{E:const}) still keeps it finite
even at $T\rightarrow 0$.
In order to eliminate the shortcoming, we replace Eq. (\ref{E:const})
by
\begin{equation}
   \sum_k
   \sum_{j=1,2,3}
    \bar{n}_{k,j}
   \sum_{i=1,2,3}
    |\psi_{ij}(k)|^2
   =(8S+4s)\,{\rm e}^{-{\mit\Delta}/k_{\rm B}T}\,.
   \label{E:newconst}
\end{equation}

\begin{figure}
\centerline
{\mbox{\psfig{figure=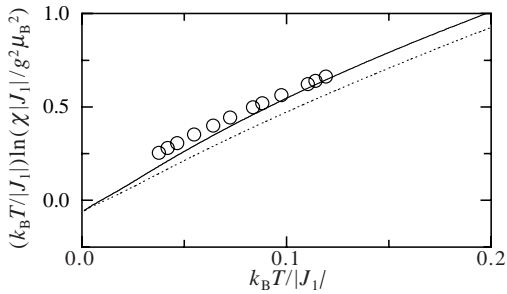,width=68mm,angle=0}}}
\vspace*{2mm}
\caption{Temperature dependences of the zero-field magnetic
         susceptibility for the parameter sets (a) (dotted line)
         and (b) (solid line).
         Experimental observations [5,9] ($\circ$) are also shown
         for reference, where we have assumed that $g=2$ [13].}
\label{F:chi}
\end{figure}

\noindent
This is quite natural modification of the theory, because the new
constraint (\ref{E:newconst}) remains the same as the authorized one
(\ref{E:const}) except for the sufficiently low-temperature region
$k_{\rm B}T\alt{\mit\Delta}$.
It is also convincing that Eq. (\ref{E:newconst}) smoothly turns into
Eq. (\ref{E:const}) as ${\mit\Delta}\rightarrow 0$.
In fact, the thus-calculated susceptibility looks reasonable in every
aspect.
In Fig. \ref{F:chi}, we make a logarithmic plot of the zero-field
susceptibility as a function of $T$ in order to elucidate its
low-temperature behavior.
Regardless of parametrization, $\chi$ exhibits an initial exponential
behavior $\propto{\rm e}^{-{\mit\Delta}/k_{\rm B}T}$ with
${\mit\Delta}\simeq 13.7\,\mbox{K}$.
The calculations are further consistent with the experimental
findings \cite{C5873,S1804}, implying that the parameter set (b)
may better describe the Mn12 cluster.
\vspace*{1mm}
\begin{figure}
\centerline
{\mbox{\psfig{figure=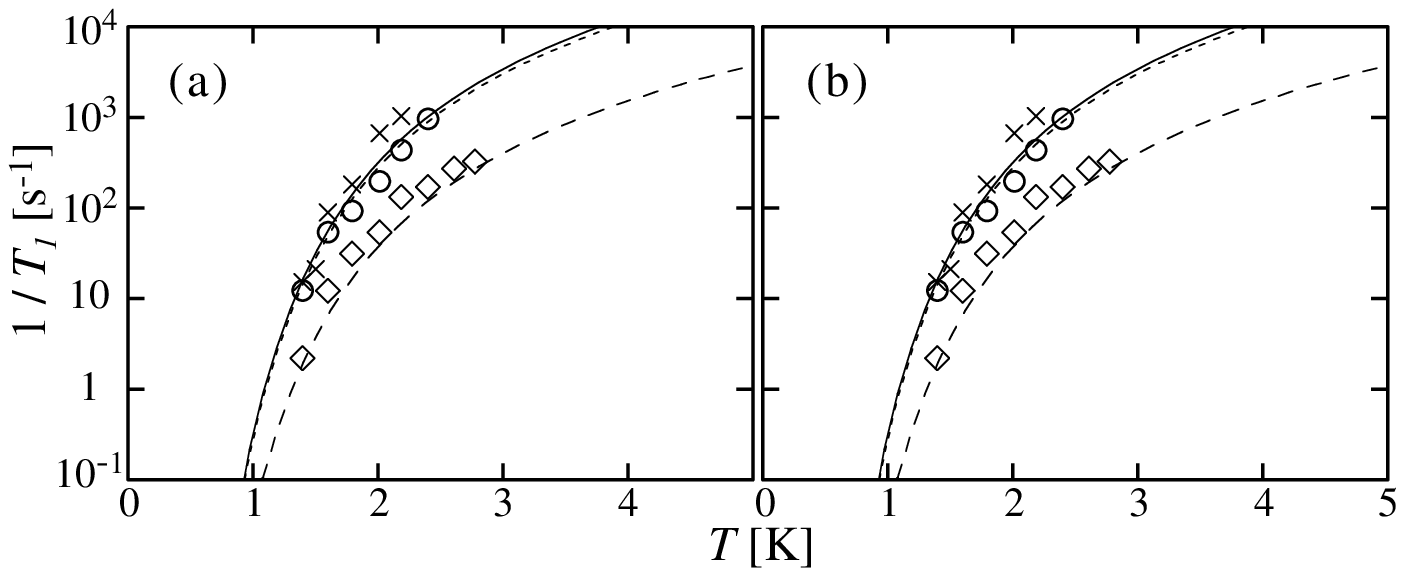,width=76mm,angle=0}}}
\vspace*{2mm}
\caption{Semilog plots of $1/T_1^{(i)}$ as a function of $T$ under no
         external field for the parameter sets (a) and (b).
         The calculations, dashed [Mn(1)], dotted [Mn(2)], and solid
         [Mn(3)] lines, are compared with experiments [19],
         $\diamond$ [Mn(1)], $\circ$ [Mn(2)], and $\times$ [Mn(3)].}
\label{F:T1T}
\vspace*{6mm}
\centerline
{\mbox{\psfig{figure=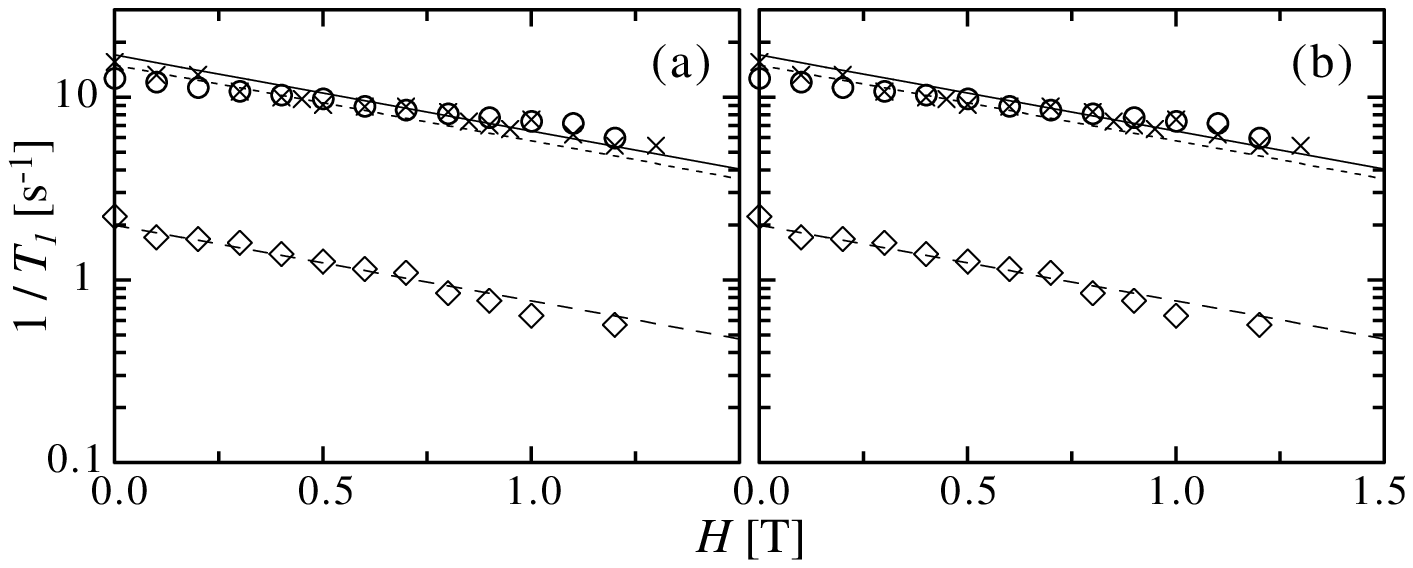,width=76mm,angle=0}}}
\vspace*{2mm}
\caption{Semilog plots of $1/T_1^{(i)}$ as a function of $H$ at
         $T=1.4\,\mbox{K}$ for the parameter sets (a) and (b).
         The calculations, dashed [Mn(1)], dotted [Mn(2)], and solid
         [Mn(3)] lines, are compared with experiments [19],
         $\diamond$ [Mn(1)], $\circ$ [Mn(2)], and $\times$ [Mn(3)].}
\label{F:T1H}
\end{figure}

   Considering the electronic-nuclear energy-conservation
requirement, the Raman process should play a leading role in the
nuclear spin-lattice relaxation.
The large zero-field energy splitting, which is associated with the
single-ion anisotropy, not only makes the direct process irrelevant
but also reduces the possibility of an odd-number-of-magnon process
being realized.
The Raman relaxation rate for the $^{55}\mbox{Mn}(i)$ nucleus is
given by
\begin{eqnarray}
   &&
   \frac{1}{T_1^{(i)}}
    =\frac{4\pi\hbar(g\mu_{\rm B}\gamma_{\rm N})^2}
          {\sum_n{\rm e}^{-E_n/k_{\rm B}T}}
     \sum_{n,m}{\rm e}^{-E_n/k_{\rm B}T}
   \nonumber \\
   &&\qquad\ \times
     \big|
      \langle m|A_i\sigma_{l,i}^z|n\rangle
     \big|^2
     \,\delta(E_m-E_n-\hbar\omega_{\rm N})\,,
\label{E:T1def}
\end{eqnarray}
where
$\sigma_{l,i}^z=s_l^z,S_l^z,\widetilde{S}_l^z$ for $i=1,2,3$,
respectively, $A_i$ is the dipolar coupling constant between the
nuclear and electronic spins on the $\mbox{Mn}(i)$ site,
$\omega_{\rm N}\equiv\gamma_{\rm N}H$ is the Larmor frequency of the
nucleus with $\gamma_{\rm N}$ being the gyromagnetic ratio, and the
summation $\sum_n$ is taken over all the electronic eigenstates
$|n\rangle$ with energy $E_n$.
Taking account of the significant difference between the electronic
and nuclear energy scales ($\hbar\omega_{\rm N}\alt 10^{-5}J$), the
relaxation rate (\ref{E:T1def}) is expressed in terms of the spin
waves as
\begin{eqnarray}
   &&
   \frac{1}{T_1^{(i)}}
    =\frac{2\hbar(g\mu_{\rm B}\gamma_{\rm N}A_i)^2}{N}
     \sum_{j=1,2,3}\sum_{\{k,k'\}}
     |\Delta\omega_j(k)|^{-1}
   \nonumber \\
   &&\qquad\ \times
     |\psi_{ij}(k)|^2 |\psi_{ij}(k')|^2
     \bar{n}_{k,j}(\bar{n}_{k',j}+1)\,,
   \label{E:T1nk}
\end{eqnarray}
where $\sum_{\{k,k'\}}$ denotes the limited summation of $(k,k')$
over $(-\pi,-\pi)$, $(-\frac{\pi}{2},-\frac{\pi}{2})$, $(0,0)$,
$(\frac{\pi}{2},\frac{\pi}{2})$, $(\frac{\pi}{2},-\frac{\pi}{2})$,
and $(-\frac{\pi}{2},\frac{\pi}{2})$, and
$\Delta\omega_j(k)=[\omega_j(k+2\pi/N)-\omega_j(k)]/(2\pi/N)$.

   We show the thus-calculated $1/T_1^{(i)}$ as functions of $T$
(Fig. \ref{F:T1T}) and $H$ (Fig. \ref{F:T1H}) in comparison with the
measurements \cite{F104401}, where the coupling constants $A_i$ are
the only adjustable parameters in reproducing both the $T$- and
$H$-dependences and have been chosen as Table \ref{T:Ai}.
Though the calculations (a) and (b) look similar, Table \ref{T:Ai}
clearly shows that the parameter assignment (b) much more reasonably
describes the Mn12 cluster.
In response to asking whether the exchange interaction $J_3$ is
ferromagnetic \cite{Z1140,C938} or antiferromagnetic
\cite{S1804,R064419}, we definitely answer that it is
antiferromagnetic.
Furthermore, we are skeptical of neglecting $J_4$ \cite{S1804,Z1140},
for which the $S=10$ ground state is much less stable \cite{R064419}
and the coupling constants are significantly underestimated in our
calculation.

   The observed temperature dependence of $1/T_1^{(i)}$, which is the
same for $i=1,2,3$, suggests that the spin dynamics of the local Mn
moments is completely correlated.
All the observations can indeed be interpreted within the Raman
relaxation process.
In the momentum summation in Eq. (\ref{E:T1nk}), the contribution
from $(j;k,k')=(1;0,0)$ ($S=10,M=9$) is predominant, while those from
$(j;|k|,|k'|)=(1,\frac{\pi}{2},\frac{\pi}{2})$ ($S=9,M=9$) at most
amount to a few percent of the total at $T=1.4\,\mbox{K}$.
The spin dynamics is thus confined within fluctuations of the total
spin moment $S=10$ of the ground state at sufficiently low
temperatures.
Although such a picture is intuitively convincing, we have to pay
attention to a recent experimental report \cite{K} that the hyperfine
field of the Mn$^{3+}$ ion is in fact anisotropic and exhibits a
predominant dipolar contribution, whereas that of the Mn$^{4+}$ ion
is isotropic and originates from the Fermi contact.
In this context, it is interesting to compare carefully the
theoretical ($A_i^{\rm th}$) and experimental ($A_i^{\rm ex}$)
findings for the coupling constants.
Assuming the set (b),
$A_1^{\rm th}\simeq 2.4A_1^{\rm ex}$,
$A_2^{\rm th}\simeq 1.7A_2^{\rm ex}$, and
$A_3^{\rm th}\simeq 1.2A_3^{\rm ex}$.
Somewhat larger deviation of the theory from the experiment for $A_1$
implies that the nuclear spin-lattice relaxation on the Mn(1) site
may not primarily be Raman active but be strongly influenced by the
surrounding Mn$^{\rm 3+}$ ions.

   The slight difference between the field dependences of
$1/T_1^{(2)}$ and $1/T_1^{(3)}$ can not be elucidated within the
present calculation, but this is probably due to the equal treatment
of the Mn(2) and Mn(3) sites, $g_2=g_3$ and $D_2=D_3$.
Considering that experimental analyses \cite{G1227,F104401} have not
yet entered into such details, our first step toward the microscopic
description of the Mn12 cluster is really successful.
We stress that the success of our spin-wave description contributes
in itself toward verifying the recently reported intramolecular spin
delocalization \cite{A064420}.
Our theory promises future investigations into various molecular
magnets over the static, dynamic, quantum, and thermal properties.

   The authors are grateful to Prof. T. Goto for fruitful discussion.
This work was supported by the Japanese Ministry of Education,
Science, and Culture and by the Sumitomo Foundation.


\begin{table}
\caption{Estimates of $(g\mu_{\rm B}\gamma_{\rm N}A_i)^2$ in the
         unit of $(\mbox{rad}\cdot\mbox{Hz})^2$ under the
         parametrizations (a) and (b) compared with experimental
         findings [19].}
\begin{tabular}{lccc}
 & Mn(1) & Mn(2) & Mn(3) \\
\noalign{\vskip 1mm}
\tableline
\noalign{\vskip 1mm}
Parameters (a) & $7.9\times 10^{17}$
               & $3.2\times 10^{18}$
               & $3.7\times 10^{18}$ \\
Parameters (b) & $6.5\times 10^{16}$
               & $2.7\times 10^{17}$
               & $3.0\times 10^{17}$ \\
Experimental   & $1.1\times 10^{16}$
               & $9.3\times 10^{16}$
               & $2.0\times 10^{17}$ \\
\end{tabular}
\label{T:Ai}
\end{table}

\widetext
\end{document}